\documentclass[final,3p]{CSP}
\usepackage{amssymb}
\usepackage{amsmath, amsthm, amssymb}
\usepackage{slashed}
\usepackage{changepage}
\usepackage{color}
\usepackage{tikz} 
\usepackage{appendix}
\usepackage{caption}
\usepackage{graphics}
\usepackage{subfigure}
\usepackage{tikz-feynman}

\usetikzlibrary{shapes,arrows,positioning,automata,backgrounds,calc,er,patterns}
\usepackage{ragged2e}
\usepackage[T1]{fontenc}
\usepackage[symbol]{footmisc}

\allowdisplaybreaks

\begin{document}

\begin{frontmatter}

\title{Lepton-pair production in di-pion  $\tau$ lepton decays}

\author[mymainaddress]{J. L. Guti\'errez Santiago\footnote[1]{jlgutierrez@fis.cinvestav.mx}}

\author[mymainaddress]{G. L\'opez Castro\footnote[2]{glopez@fis.cinvestav.mx}}
\author[mymainaddress]{P. Roig\footnote[3]{proig@fis.cinvestav.mx}}

\address[mymainaddress]{Departamento de F\'isica, Centro de Investigaci\'on y de Estudios Avanzados del Instituto Polit\'ecnico Nacional,\\ Apdo. Postal 14-740, 07000 Ciudad de M\'exico, M\'exico.}

\begin{abstract}\justify\rm
\begin{adjustwidth}{2cm}{2cm}{\itshape\textbf{Abstract:}}
We  study the $\tau^- \to\nu_{\tau} \pi^{-}\pi^{0}\ell^{+}\ell^{-}$ ($\ell=e,\,\mu$) decays, which are $O(\alpha^2)$-suppressed with respect to the dominant di-pion tau decay channel. Both the inner-bremsstrahlung and the structure- (and model-)dependent contributions are considered. In the $\ell=e$ case, structure-dependent effects are $\mathcal{O}(1\%)$ in the decay rate, yielding a clean prediction of its branching ratio, $2.3\times10^{-5}$, measurable with BaBar or Belle(-II) data. For $\ell=\mu$, both contributions 
have similar magnitude and we get a branching fraction of $(1.6\pm0.3)\times10^{-7}$, reachable by the end of Belle-II operation. These decays allow to study the dynamics of strong interactions with simultaneous weak and electromagnetic probes; their knowledge will contribute to reducing backgrounds in lepton flavor/number violating searches. 
\end{adjustwidth}
\end{abstract}

\end{frontmatter}

\section{Introduction}

Among the different charged lepton flavours, the $\tau$ lepton has the largest mass, enough to produce a variety of hadronic states which provide an ideal environment to study the dynamics of the strong interactions as far as phase space allows. In this sense, $\tau$ lepton decays involve energy domains where the resonance degrees of freedom become relevant. By means of semileptonic $\tau$ decays, we can study the hadronization of the weak charged currents and use the resulting hadronic vertices either to test the fundamental parameters of the Standard Model (SM) or to understand the properties of Quantum Chromodynamics 
and the electroweak sectors in a clean way \cite{Davier, Pich:2013lsa}.

This article studies the semileptonic five-body decay $\tau^{-} \to \nu_{\tau}\pi^{-}\pi^{0}\ell^{+}\ell^{-}$ with the  lepton pair ($\ell=e$ or $\mu$) produced via  a virtual photon. The corresponding radiative case $\tau^-\to \nu_{\tau}\pi^-\pi^0\gamma$ has been analyzed using the Resonance Chiral Theory (RChT) \cite{RChT1, RChT2, Miranda:2020wdg} and vector meson dominance  \cite{FloresTlalpa:2005fz, FloresBaez:2006gf} approaches  to describe  vector and axial-vector form factors involved in the $W^- \to \pi^-\pi^0\gamma$ vertex  (see also ref. \cite{Jegerlehner:2011ti}, which 
includes isospin-breaking and electromagnetic radiative corrections according to the Hidden Local Symmetry model, last updated in ref. \cite{Benayoun:2019zwh}). Interestingly, this weak vertex  involves also the interplay with strong and electromagnetic interactions. From a more practical point of view, the study of this decay is useful because it may pollute searches for processes involving lepton flavor violation in the charged sector %
or 
lepton number violation. In addition, it could serve to verify the radiative corrections used in the contribution of the hadronic vacuum polarization (HVP) entering the anomalous magnetic moment of the muon ($a_\mu$) obtained using  hadronic $\tau$ decays data \cite{Alemany:1997tn, Cirigliano:2001er, Cirigliano, FloresBaez:2006gf, FloresTlalpa:2006gs,Davier:2009ag, Jegerlehner:2011ti, Davier:2010nc, Davier:2013sfa, Miranda:2020wdg}.

Specifically, the decay $\tau^{-} \to \nu_{\tau}\pi^{-}\pi^{0}\ell^{+}\ell^{-}$, with $\pi^-$ misidentified as muon and undetected $\pi^0$, may be an important background that can mimic the signal in searches for lepton flavour violation (LFV) processes of the form $\tau^{-}\to \ell^{(\prime)-}\ell^{+}\ell^{-}$
. Currently the branching ratios of these LFV decay modes have upper bounds of $O(10^{-8})$ \cite{PDG}, while their SM predictions are unmeasurably small \cite{Hernandez-Tome:2018fbq, Blackstone:2019njl}. These decays (for $\ell=\mu$) can also be misidentified as lepton number violating processes of the type $\tau^- \to \nu_\tau \mu^- \mu^- \pi^+$ \cite{Castro:2012gi}. To avoid these decays polluting new physics searches, it will be most useful to include them in the Monte Carlo Generator TAUOLA \cite{Shekhovtsova:2012ra}, where other tau decay modes including $\ell^+\ell^-$ pairs \cite{Adolfo, Flores-Tlalpa:2015vga} have recently been incorporated \cite{Antropov:2019ald}.

Ref. \cite{Alemany:1997tn} first took advantage of the clean LEP tau data samples to evaluate $a_\mu^{HVP,LO}$ using tau data.  At the level of precision attained in the last twenty years (see ref. \cite{Aoyama:2020ynm} and references therein), one requires computing the (model-dependent) electromagnetic and isospin breaking corrections relating $\sigma(e^+e^-\to {\rm hadrons})$ in the isovector channel to the hadronic tau decay distributions, which were taken into account in subsequent evaluations \cite{Cirigliano:2001er, Cirigliano, FloresBaez:2006gf, FloresTlalpa:2006gs,Davier:2009ag, Jegerlehner:2011ti, Davier:2010nc, Davier:2013sfa, Miranda:2020wdg}. Particularly, refs. \cite{Cirigliano, FloresBaez:2006gf, Miranda:2020wdg} highlighted that different observables of $\tau^-\to\nu_\tau\pi^-\pi^0\gamma$ decays for a not-so-low cut on photon energies (so that the inner bremsstrahlung (IB) part does not saturate the observables) can reduce substantially the model-dependent error in $\tau$-based evaluations of $a_\mu^{HVP,LO}$. We hope that the analysis presented in this work of the  $\tau^-\to\nu_\tau\pi^-\pi^0\ell^+\ell^-$ decays can be helpful in reducing the error of $a_\mu^{HVP,LO}$ using tau data. We will find that the $\ell=\mu$ case is promising in this respect, as the model-dependent contributions are of the same size of the IB part. Conversely, its low branching ratio, $\sim10^{-7}$ will challenge the Belle-II analysis \cite{Kou:2018nap}. For the $\ell=e$ case the situation will be the opposite, with a $\sim10^{-5}$ branching ratio (measurable already with BaBar or Belle data), that has little ($\mathcal{O}(\%)$) model-dependence.

The IB part is model-independent \cite{Low:1958sn}, while the structure-dependent part is not.  The vector and axial-vector form factors entering the latter can be computed using the RChT framework. Including operators contributing -upon integrating resonances out- to the  $\mathcal{O}(p^4)$ Chiral Perturbation Theory (ChPT) \cite{Gasser:1983yg} low-energy couplings --as in ref. \cite{Cirigliano}-- all free parameters are related to the pion decay constant (after applying short-distance QCD constraints), which results in controlled ($\mathcal{O}(20\%)$) errors for our prediction in the $\ell=\mu$ mode. In the $\ell=e$ case, structure-dependent effects are as small as (uncomputed) one-loop QED corrections, which set the size of our corresponding uncertainty.

The paper is organized as follows: we start with a very short review of the radiative process $\tau^- \to \nu_{\tau}\pi^{-}\pi^{0}\gamma$ in order to introduce our conventions and recall the main features of the process under study. In section \ref{T5C} we describe the amplitude for lepton pair production $\tau^- \to \nu_{\tau}\pi^-\pi^0\ell^+\ell^-$.  Section \ref{ssd} deals with both, vector and axial-vector, structure-dependent amplitudes. We derive the corresponding basis for the relevant (vector) case at this order in the chiral expansion in section \ref{sectionvector}, relegating the axial-vector structure (which only appears at the next order) to section \ref{axc}
. 
The branching ratio and the $\ell^+\ell^-$ invariant mass spectrum for both channels are predicted in section \ref{BRIMS}. Finally, we provide our conclusions in section \ref{Conclusions}.

\section{The radiative $\tau^{-} \to \nu_\tau\pi^{-}\pi^{0} \gamma$ decay}

The  $\tau^- \to \nu_{\tau}\pi^-\pi^0$ decay is the dominant channel among tau decays. The precise measurement of the di-pion mass spectrum allows to extract the weak pion form factor, which can be compared to the electromagnetic pion form factor measured in electron-positron collisions via the Conserved Vector Current (CVC)  hypothesis. This same spectrum is useful to extract, in a clean way,  information  on the tower of vector resonance parameters that are produced in the hadronization of the isovector current. On the other hand the radiative di-pion tau lepton decay, $\tau^- \to \nu_{\tau}\pi^-\pi^0\gamma$, provides additional information on the hadronization of the weak current and it is necessary to account for the di-pion observables at the few percent level. Finally, the lepton-pair production induced by the virtual photon in the radiative decay, namely  $\tau^- \to \nu_{\tau}\pi^-\pi^0\gamma^*(\to \ell^+\ell^-)$, serves to further scrutinize the hadronization of the weak current in an extended kinematical domain. As previously pointed out, it allows also to quantify one of the sources of background in searches of three charged lepton flavor violating decays of tau leptons.

In order to fix our conventions in the study of lepton-pair production, let us first consider the radiative di-pion decays of the tau lepton. 
The matrix element for the process $\tau^{-}(P)\to \nu_\tau(q) \pi^{-}(p_{-})\pi^{0}(p_{0})\gamma(k)$, has the general structure \cite{Cirigliano}:
\begin{eqnarray}\label{ACirigliano}
    \mathcal{M}(\gamma) &=& eG_{F}V_{ud}^{*}\epsilon^{*\mu}(k)\Bigg[ \frac{(p_{0}-p_{-})_{\nu}f_{+}(t)}{2P\cdot k}\overline{u}(q)\gamma^{\nu}(1-\gamma_{5})\left(\slashed{P} - \slashed{k} + M_{\tau} \right)\gamma_{\mu}u_{\tau}(P) 
    .\nonumber\\
    &+& 
    \left( V_{\mu\nu} - A_{\mu\nu} \right)\overline{u}(q)\gamma^{\nu}(1-\gamma_{5})u_{\tau}(P) \Bigg] .
\end{eqnarray}
Here, $G_F$ is the Fermi coupling constant, $V_{ud}$ the Cabibbo-Kobayashi-Maskawa quark mixing matrix element, $e$ is the magnitude of the electron charge and $\epsilon^{*\mu}(k)$ the photon polarization four-vector.
The first term in Eq. (\ref{ACirigliano}) corresponds to  the photon emission off the tau lepton. It is described in terms of the 
charged pion vector form factor, defined as 
\begin{equation}
\langle \pi^{-}\pi^{} |\overline{d}\gamma^{\mu} u |0\rangle = \sqrt{2}f_{+}(t) \left(p_{-}-p_{0} \right)^{\mu}, \nonumber     
\end{equation}
 {with $t=(p_{-}+p_{0})^2$ the invariant-mass of the di-pion system. Later, in our numerical analysis we will use $f_+(x)$ as derived in Refs. \cite{Dumm:2013zh, Gonzalez-Solis:2019iod}.

The second term in Eq. (\ref{ACirigliano}) contains the (structure-dependent) vector $V_{\mu\nu}$ and axial-vector $A_{\mu\nu}$ components. They describe the hadronization of the weak current involving an additional photon: 
\begin{equation}\label{Wvertex}
    W^{-}(P-q)\to \pi^{-}(p_{-})\pi^{0}(p_{0})\gamma(k). \  
\end{equation}
According to Low's soft-photon theorem \cite{Low:1958sn}, the leading terms (LO) of the radiative amplitude in the photon energy expansion are fixed in terms of the non-radiative amplitude and the gauge invariance requirement of the total amplitude (equivalently, $k^{\mu}V_{\mu\nu} =f_{+}(t)(p_{-}-p_{0})_\nu$). This leads to ($t'=(p_-+p_0+k)^2=t+2(p_-+p_0)\cdot k$)
\begin{equation}
    V_{\mu\nu}^{LO}=f_{+}(t^{\prime})\frac{p_{- \mu}}{p_{-}\cdot k}\left(p_{-}+k-p_{0} \right)_{\nu} - f_{+}(t^{\prime})g_{\mu\nu} + \frac{f_{+}(t^{\prime})-f_{+}(t)}{(p_{-}+p_{0})\cdot k}\left( p_{-}+p_{0}\right)_{\mu}\left( p_{0}-p_{-}\right)_{\nu}. \nonumber \\ 
\end{equation}
\\
\noindent In addition, the full vector contribution to the radiative amplitude contains model-dependent gauge-invariant terms of $O(k)$, such that $V_{\mu\nu}=V_{\mu\nu}^{LO}+\widehat{V}_{\mu\nu}$. Low's soft-photon theorem \cite{Low:1958sn} is manifestly satisfied since \cite{Cirigliano}:
\begin{eqnarray}
V_{\mu\nu}^{LO} &=& f_{+}(t)\frac{p_{-\mu}}{p_{-}\cdot k}(p_{-}-p_{0})_{\nu} + f_{+}(t)\left(  \frac{p_{-\mu}k_{\nu}}{p_{-}\cdot k} - g_{\mu\nu} \right) \nonumber \\
&+& 2 \frac{df_{+}(t)}{dt}\left(\frac{p_{-\mu}p_{0}\cdot k}{p_{-}\cdot k} - p_{0\mu}  \right)(p_{-}-p_{0})_{\nu} + \mathcal{O}(k).\nonumber
\end{eqnarray}

Since axial-vector contributions $A_{\mu\nu}$ are not present in the non-radiative amplitude, they start at $O(k)$ in the photon energy expansion. Thus, they are model-dependent  and must be manifestly gauge-invariant: $k^{\mu}A_{\mu\nu}=0$.  
Structure-dependent vector and axial-vector contributions to $\tau^{-} \to \nu_\tau\pi^{-} \pi^{0} \ell^{+}\ell^{-}$ decays, which reproduce the corresponding terms of the radiative decay in the case of the real photon ($k^2=0$), are considered in section \ref{ssd}.
 
\section{The $\tau^{-} \to \nu_\tau\pi^{-} \pi^{0} \ell^{+}\ell^{-}$ amplitude: 
model-independent contribution} \label{T5C}

In this section we consider the leading terms (in the virtual photon momentum expansion) of the amplitude for lepton-pair production in the di-pion tau lepton decay. As stated before, they depend only upon the form factors and electromagnetic properties of the particles involved in the non-radiative amplitude and are fixed from the gauge invariance requirement. The contributions to this part of the amplitude are given by the diagrams shown in Fig. \ref{FD}. 

\begin{center}
\begin{figure}[h]
    \centering
    \subfigure[]{%
        \begin{tikzpicture}
            \begin{feynman}
                \vertex (a) {\(\tau^{-}\)};
                \vertex [right = of a] (b);
                \vertex [above = 0.7cm of b] (c);
                \vertex [above left = 0.7cm of c] (d) {\(\ell^{+}\)};
                \vertex [above right = 0.7cm of c] (e) {\(\ell^{-}\)};
                \vertex [small, right = of b] (f);
                \vertex [above right=of f] (g) {\(\nu_{\tau}\)};
                \vertex [small, below right=of f, dot] (h){};
                \vertex [small, above right=of h] (p0) {\(\pi^{-}\)};
                \vertex [small, below right=of h] (pm) {\(\pi^{0}\)};
                \diagram* {
                (a) -- [fermion] (b) -- [fermion] (f) -- [fermion] (g),
                (b) -- [photon, edge label'=\(\gamma\)] (c),
                (c) -- [anti fermion] (d),
                (c) -- [fermion] (e),
                (f) -- [boson, edge label=\(W^{-}\), ] (h),
                (h) -- [scalar] (p0),
                (h) -- [scalar] (pm),
                };
            \end{feynman}
        \end{tikzpicture}
        \label{D1}
    }%
    \subfigure[]{%
        \begin{tikzpicture}
            \begin{feynman}
                \vertex (a) {\(\tau^{-}\)};
                \vertex [right = of a] (b);
                \vertex [above right = of b] (c) {\(\nu_{\tau}\)};
                \vertex [right = of b] (d);
                \vertex [below right = of d] (p0) {\(\pi^{0}\)};
                \vertex [small, above right = of d, dot] (e) {};
                \vertex [small, above right = of e] (pm) {\(\pi^{-}\)};
                \vertex [below right = 0.7cm of e] (f);
                \vertex [above right = 0.7cm of f] (g) {\(\ell^{+}\)};
                \vertex [below right = 0.7cm of f] (h) {\(\ell^{-}\)};
                \diagram* {
                (a) -- [fermion] (b) -- [boson, edge label=\(W^{-}\), near end] (d),
                (b) -- [fermion] (c),
                (d) -- [scalar] (p0),
                (d) -- [scalar] (e),
                (e) -- [scalar] (pm),
                (e) -- [photon, edge label=\(\gamma\)] (f),
                (f) -- [anti fermion] (g),
                (f) -- [fermion] (h),
                };
            \end{feynman}
        \end{tikzpicture}
        \label{D2}
    }%
    \subfigure[]{
        \begin{tikzpicture}
            \begin{feynman}
                \vertex (a) {\(\tau^{-}\)};
                \vertex [right = of a] (b);
                \vertex [above right = of b] (c) {\(\nu_{\tau}\)};
                \vertex [right = of b, dot] (d) {};
                \vertex [below right = of d] (p0) {\(\pi^{0}\)};
                \vertex [above right = of d] (pm) {\(\pi^{-}\)};
                \vertex [small, right = of d] (e);
                \vertex [above right = 0.6cm of e] (f) {\(\ell^{+}\)};
                \vertex [below right = 0.6cm of e] (g) {\(\ell^{-}\)};
                \diagram* {
                (a) -- [fermion] (b) -- [fermion] (c),
                (b) -- [boson, edge label=\(W^{-}\), near end] (d),
                (d) -- [photon, edge label=\(\gamma\)] (e),
                (d) -- [scalar] (p0),
                (d) -- [scalar] (pm),
                (e) -- [anti fermion] (f),
                (e) -- [fermion] (g),
                };
            \end{feynman}
        \end{tikzpicture}
        \label{D3}
    }
\caption[]{Feynman diagrams contributing to the $\tau^{-} \to \nu_\tau \pi^{-} \pi ^{0}\ell^{+}\ell^{-} $ decay: \subref{D1} Inner Bremsstrahlung (IB) off the tau lepton; \subref{D2} IB off the $\pi^{-}$ meson, \subref{D3}  contact term.}
\label{FD}
\end{figure}
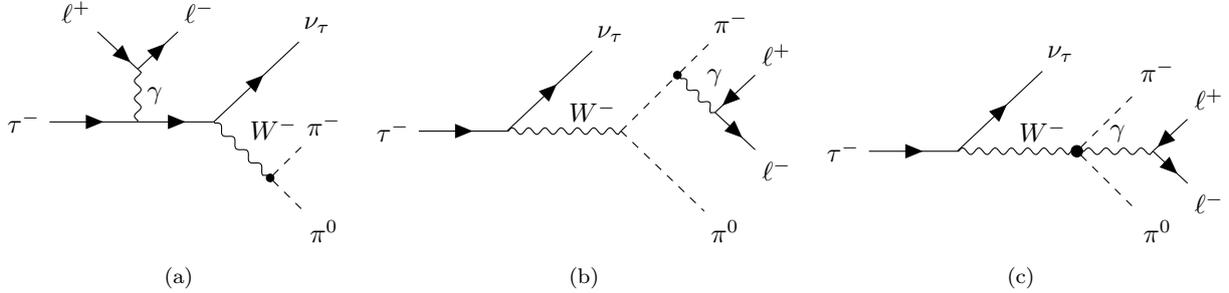
\end{center}

The matrix element for the decay $\tau^{-}(P)\to \nu_\tau(q) \pi^{-}(p_{-})\pi^{0}(p_{0})\ell^{+}(p_{\ell^+})\ell^{-}(p_{\ell^-})$ has a similar structure to the radiative amplitude:
\begin{eqnarray}\label{GS}
\mathcal{M} (\ell\ell)&=& \frac{e^2}{k^2}G_{F}V^{*}_{ud}l^{\mu}\Bigg\{\frac{\left(p_{-}-p_{0} \right)_{\nu}}{k^{2}-2P\cdot k} f_{+}(t)\overline{u}_{\nu}(q)\gamma^{\nu}(1-\gamma_{5}) \left(\slashed{P}-\slashed{k}+M_\tau\right)\gamma_{\mu}u_{\tau}(P) 
\nonumber \\
&\ \ \ \  \ \ \ \ +& 
\left( V_{\mu\nu} - A_{\mu\nu} \right)L^{\nu} \Bigg\},
\end{eqnarray}

\noindent where we have defined the weak $L^{\nu} = \overline{u}_{\nu}(q)\gamma^{\nu}(1-\gamma_{5})u_{\tau}(P)$ and electromagnetic  $l^{\mu} = \overline{u}(p_{\ell^-})\gamma^{\mu}v(p_{\ell^+})$ leptonic currents. Owing to  Dirac's equation and $k=p_{\ell^+}+p_{\ell^-}$, we have  $l^{\mu}k_{\mu}=0$. The variable $t$ is still defined as the invariant mass of the two-pion system, however for virtual photons $t' \equiv (P-q)^2=t+2(p_-+p_0)\cdot k+k^2$ . 

As in the case of the radiative amplitude, we split the vector contribution into two terms: 
\begin{equation}\label{vector}
V_{\mu\nu} =V_{\mu\nu}^{LO}+\widehat{V}_{\mu\nu}\ .
\end{equation} 
The leading order LO (model-independent) contribution is fixed from the diagrams  in Fig. \ref{FD} and the gauge-invariance requirement. One gets:

\begin{center}
\begin{eqnarray}\label{Vmunu}
V_{\mu\nu}^{LO} &=& f_{+}(t^{\prime})\frac{2p_{-\mu}+k_{\mu}}{2p_{-}\cdot k + k^2}\left(p_{-}+k-p_{0}\right)_{\nu} - f_{+}(t^{\prime})g_{\mu\nu} \nonumber \\
&+& \left( \frac{f_{+}(t^{\prime})-f_{+}(t)}{2(p_{0}+p_{-})\cdot k + k^2} \right) \left( 2 \left(p_{-}+p_{0}\right)_{\mu} +k_{\mu}\right) \left( p_{-}-p_{0}\right)_{\nu}.  \\ \nonumber 
\end{eqnarray}
\end{center}
\noindent The terms proportional to $k_{\mu}$  can be dropped ($k^{\mu}l_{\mu}=0$) in the previous expression owing to the conservation of the electromagnetic current. Note also the presence of $k^2$ terms appearing in denominators of  the propagators of charged particles.

This model-independent amplitude, also known as the inner bremsstrahlung term in this paper},  is the leading contribution at low photon momenta given the off-shell propagators of charged particles, and the enhancement factor provided  by  the  photon pole propagator. This feature makes model-independent contributions more important for observables in $e^+e^-$, which has a lower threshold than for muon-pair production. On the other hand,  model or structure-dependent contributions, described by $\widehat{V}_{\mu\nu}-A_{\mu\nu}$ terms, start at $O(k)$ and may become important for large photon momenta. The effects of model-dependent contributions become more visible in the $\mu^+\mu^-$ pair production, allowing to explore the rich dynamics of strong interactions in the intermediate energy regime. We will consider the later contributions in the following section.

\section{Structure-dependent contributions}\label{ssd}

The structure-dependent piece of vector contributions to the amplitude, $\widehat{V}_{\mu\nu}$, as well as axial-vector contributions $A_{\mu\nu}$ are computed in the framework of resonance chiral theory. They are analogous to similar amplitudes in the radiative di-pion decays of tau leptons, but now they involve a virtual photon ($k^2=(p_{\ell^+}+p_{\ell^-})^2$).

The hadronic vertex $W^*(W)\to \pi^-(p_-)\pi^0(p_0)\gamma^*(k)$ with two virtual gauge bosons (characterized by $W^2=t'=(p_-+p_0+k)^2\not = m_W^2,\ k^2=(p_{\ell^+}+p_{\ell^-})^2\not = 0$) can be parameterized in terms of two different sets of form factors, vector and axial-vector. In addition to the squared momenta of these virtual bosons ($t',k^2$), the form factors can depend upon two independent kinematical variables which can be taken as ($t,s_-$) or ($t, s_0$), where $t=(p_-+p_0)^2$ is the invariant mass of the di-pion system and $s_{-,0}\equiv (p_{-,0}+k)^2$, or equivalently $p_-\cdot k$ or $p_0\cdot k$. Once we chose ($t,t',k^2$) as relevant variables, either $p_-\cdot k$ or $p_0\cdot k$ can be chosen as the remaining kinematical scalar to describe the hadronic vertex. 

In this section we discuss the Lorentz structure of the vector and axial-vector contributions to the hadronic vertex. We first compute the expressions for the vector form factors within RChT and end this subsection by discussing the short-distance QCD constraints on the resonance couplings. In the second part we comment on the axial-vector contributions. Since they play a subleading role numerically (as checked in the real photon case \cite{FloresTlalpa:2005fz, FloresBaez:2006gf, Miranda:2020wdg})
, we will consider only the terms arising  from the axial anomaly and the Wess-Zumino contributions for the case of real photons as computed in \cite{Cirigliano}. 

\subsection{Structure-dependent vector contributions}\label{sectionvector}

As stated before, the most general form of the structure-dependent part of the vector contributions to the hadronic vertex can be built out of the metric tensor $g_{\mu\nu}$ and the three independent momenta $p_-,\ p_0$ and $k$ by imposing gauge-invariance. Including the leading order terms (\ref{Vmunu}), we get :
\begin{eqnarray}\label{VP}
V^{\mu\nu} &=& v_{1}\left( p_-\cdot k g^{\mu\nu} - p_{-}^{\mu}k^{\nu} \right) + v_{2}\left( p_0\cdot k g^{\mu\nu} - p_{0}^{\mu}k^{\nu} \right) + v_{3}\left( p_0\cdot k p_{-}^{\mu} - (p_-\cdot kp_{0}^{\mu} \right) p_{-}^{\nu} \nonumber\\
&+& v_{4}\left( p_0\cdot kp_{-}^{\mu} - p_-\cdot k p_{0}^{\mu} \right)\left(p_{-}+p_{0}+k\right)^{\nu} + v_{5} \left( k^2g^{\mu\nu}-k^{\mu}k^{\nu}\right) +v_{6} \left( k^2p_{-}^{\mu}-p_-\cdot k k^{\mu}\right) p_{0}^{\nu} \nonumber \\
&+& v_{7} \left( p_0\cdot k k^{\mu} - k^2 p_{0}^{\mu} \right)p_{-}^{\nu} -f_{+}(t^{\prime})g^{\mu\nu} + \frac{f_{+}(t^{\prime})-f_{+}(t)}{2 \left( p_{0}+p_{-} \right)\cdot k+k^2}\left( 2(p_{0}+p_{-})+k^{\mu} \right)^{\mu}\left( p_{0}-p_{-} \right)^{\nu} \nonumber \\
&+& \frac{f_{+}(t^{\prime})}{2k \cdot p_{-}+k^2} \left( 2p^{\mu}_{-}+k^{\mu}\right) \left( p_{-}+k-p_{0} \right)^{\nu}\ .
\end{eqnarray} 
The Lorentz-invariant form factors $v_{1,\cdots ,7}$ depend in general upon the four invariant variables discussed above. They encode the information about the dynamics of the strong, weak and electromagnetic interactions involved in the $W^*\pi^-\pi^0\gamma^*$ vertex. They are the coefficients of (explicitly gauge-invariant) Lorentz vector tensor structures. In the case of a real photon, $v_{5,6,7}$ do not contribute to the amplitude given that $k^2=0$ and $\epsilon \cdot k=0$ and one recovers the results of Ref. \cite{Cirigliano} for the radiative amplitude, as it should be. Although terms proportional to $k^{\mu}$ do not contribute to physical results owing to current conservation, we keep them because explicit calculations of the hadronic vertex gives rise to such structures and in order to exhibit explicitly gauge invariance.

Next, we compute the different Feynman diagrams appearing in Figure \ref{CFD} 
within the RChT framework  \cite{RChT1, RChT2}, which ensures the low-energy behaviour of ChPT \cite{Gasser:1983yg} and includes resonances as dynamical degrees of freedom upon their approximate $U(3)$ flavor symmetry. 

Besides the kinetic terms for the resonances (that we do not quote), we have used the interaction  Lagrangian given by (see ref. \cite{RChT1} for further details)
\begin{eqnarray}
\label{InteractionLagrangian}
\mathcal{L}_{2V}&=&\frac{F_{V}}{2\sqrt{2}}Tr\left( \widetilde{V}_{\mu\nu}f_{+}^{\mu\nu} \right) + i\frac{G_{V}}{\sqrt{2}}Tr\left( \widetilde{V}_{\mu\nu}u^{\mu}u^{\nu} \right),\nonumber\\
\mathcal{L}_{2A}&=&\frac{F_{A}}{2\sqrt{2}}Tr\left( \widetilde{A}_{\mu\nu}f_{-}^{\mu\nu} \right), \nonumber
\end{eqnarray}
\noindent where $Tr$ stands for a trace in flavor space. Resonance fields are represented by the antisymmetric tensors \cite{ATF1, ATF2} $\widetilde{V}_{\mu\nu}$ and $\widetilde{A}_{\mu\nu}$ (not to be confused with the tensor contributions to the decay amplitude defined in Eq. (\ref{ACirigliano})), the coupling to the weak charged $V-A$ current proceeds through the $f_{\pm}^{\mu\nu}$ tensors, and the $u^\mu$ tensors couple the resonances to either the vector part of the $W$ boson or (derivatively) to pion fields. 
The coupling constants of resonances  $F_V,G_V$ and $F_A$ can be fixed from short-distance constraints in terms of $f = F_{\pi} \sim 92$ MeV.
Short-distance QCD constraints on the spin-one correlators \cite{RChT1, RChT2, Weinberg:1967kj} predict the former in terms of the latter as $F_V = \sqrt{2}f, \ G_V = f/\sqrt{2},\  F_A = f$. The associated uncertainties are discussed below.

In figure \ref{CFD} we show the Feynman diagrams that contribute to  the vector tensor amplitudes in Eq. (\ref{VP}). 
For convenience and later comparison, we quote here the results in the real photon case \cite{Cirigliano} 
and defer to the end of this subsection  their generalization for a virtual photon:
\begin{eqnarray}\label{firstap}
v_{1} &=& \frac{F_{V}G_{V}}{f^2m_{\rho}^2}\left[ 2 + 2m_{\rho}^2D^{-1}_{\rho}(t^{\prime}) + tD^{-1}_{\rho}(t) + tm_{\rho}^2 D_{\rho}^{-1}(t)D_{\rho}^{-1}(t^{\prime}) \right] \nonumber \\
&\ \ \ \ \ +& \frac{F_{V}^2}{2f^2m_{\rho}^2}\left[ -1-m_{\rho}^2D_{\rho}^{-1}(t^{\prime}) + t^{\prime}D_{\rho}^{-1}(t^{\prime}) \right] + \frac{F_{A}^2}{f^2m_{a_1}^2}\left[ m_{a_1}^2 - m_{\pi}^2 + \frac{t}{2} \right]D_{a_1}^{-1}\left[ (p_{-}+k)^2 \right], \nonumber\\
v_{2} &=& \frac{F_{V}G_{V}t}{f^2m_{\rho}^2}\left[ -D_{\rho}^{-1}(t) - m_{\rho}^2D_{\rho}^{-1}(t)D_{\rho}^{-1}(t^{\prime}) \right] + \frac{F_{V}^2}{2f^2m_{\rho}^2} \left[ -1 - m_{\rho}^2D_{\rho}^{-1}(t^{\prime}) - t^{\prime}D_{\rho}^{-1}(t^{\prime}) \right] \nonumber \\
&\ \ \ \ \ +& \frac{F_{A}^2}{f^2m_{a_1}^2}\left[ m_{a_1}^2 - m_{\pi}^2 - p_{-}\cdot k \right] D_{a_1}^{-1}\left[ (p_{-}+k)^2 \right],\nonumber\\
v_{3} &=& \frac{F_{A}^2}{f^2m_{a_1}^2} D_{a_1}^{-1}\left[ (p_{-}+k)^2 \right],\nonumber\\
v_{4} &=& -\frac{2F_{V}G_{V}}{f^2}D_{\rho}^{-1}(t)D_{\rho}^{-1}(t^{\prime}) + \frac{F_{V}^2}{f^2m_{\rho}^2}D_{\rho}^{-1}(t^{\prime}),
\end{eqnarray}
\noindent where the propagators are given by
\begin{eqnarray}
D_{\rho}(s) &=& m_{\rho}^2 - s - im_{\rho}\Gamma_{\rho}(s),\nonumber\\
D_{a_1}(s) &=& m_{a_1}^2 - s - im_{a_1}\Gamma_{a_1}(s).
\end{eqnarray}
The off-shell widths of the $\rho(770)$ and $a_1(1260)$ mesons that appear in the above expressions  and are used in this paper, are obtained within RChT. In the first case it includes the $\pi\pi$ and $K\bar{K}$ cuts \cite{Guerrero:1997ku, GomezDumm:2000fz} and in the second case the $3\pi$ \cite{Dumm:2009va, Nugent:2013hxa} and $K\bar{K}\pi$ \cite{Dumm:2009kj} cuts.
\begin{center}
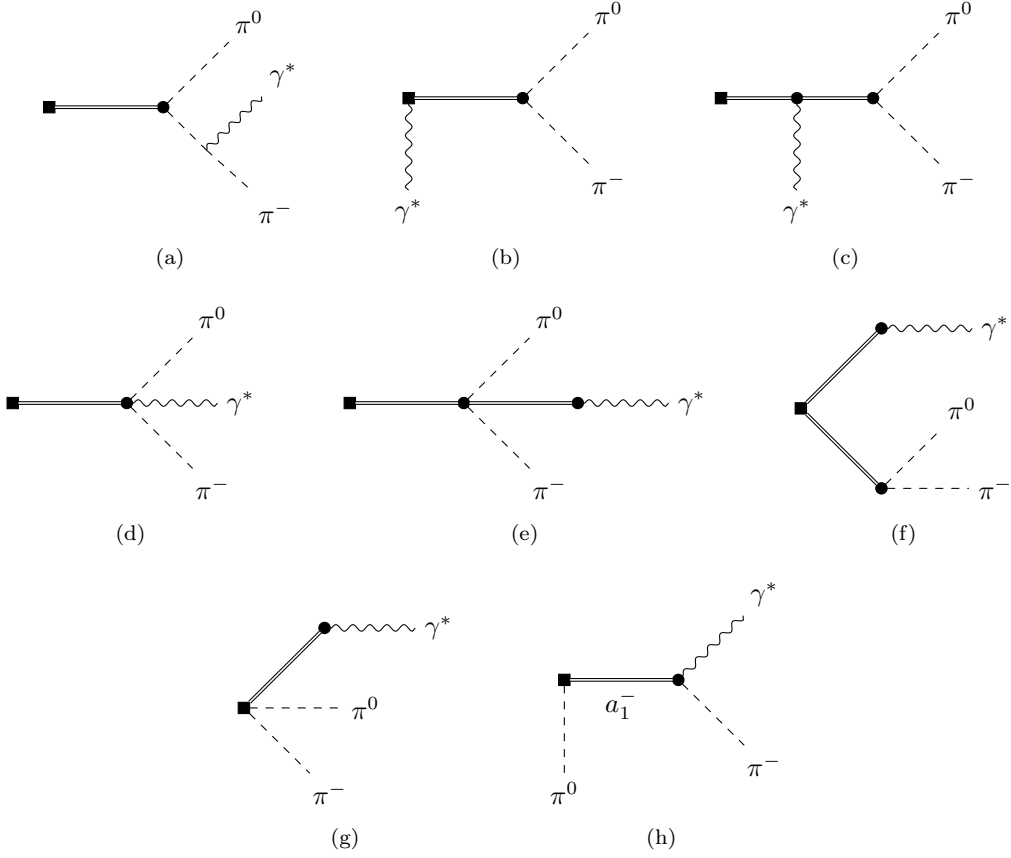
\begin{figure}[h]
    \centering
    \subfigure[]{
        \begin{tikzpicture}
            \begin{feynman}
                \vertex [square dot] (a) {};
                \vertex [right=of a,dot] (b){};
                \vertex [above right=1.6cm of b] (p1) {\(\pi^{0}\)};
                \vertex [below right=0.8cm of b] (c);
                \vertex [below right=0.8cm of c] (p2) {\(\pi^{-}\)};
                \vertex [above right=1cm of c] (g) {\(\gamma^*\)};
                \diagram* {
                (a) -- [double] (b) -- [scalar] (p1),
                (b) -- [scalar] (c),
                (c) -- [scalar] (p2),
                (c) -- [photon] (g),
                };
            \end{feynman}
        \end{tikzpicture}
        \label{RCT1}
    }
    \quad \quad 
    \subfigure[]{
        \begin{tikzpicture}
            \begin{feynman}
                \vertex [square dot] (a) {};
                \vertex [below=of a] (b) {\(\gamma^*\)};
                \vertex [right=of a, dot] (c){};
                \vertex [above right=1.6cm of c] (p1) {\(\pi^{0}\)};
                \vertex [below right=1.6cm of c] (p2) {\(\pi^{-}\)};
                \diagram* {
                (a) -- [photon] (b),
                (a) -- [double] (c),
                (c) -- [scalar] (p1),
                (c) -- [scalar] (p2),
                };
            \end{feynman}
        \end{tikzpicture}
        \label{RCT2}
    }
    \quad \quad 
    \subfigure[]{
        \begin{tikzpicture}
            \begin{feynman}
                \vertex [square dot] (a) {};
                \vertex [right=1cm of a,dot] (b){};
                \vertex [below=of b] (c){\(\gamma^*\)};
                \vertex [right=1cm of b, dot] (d) {};
                \vertex [above right=1.6cm of d] (p1){\(\pi^{0}\)};
                \vertex [below right=1.6cm of d] (p2) {\(\pi^{-}\)};
                \diagram* {
                (a) -- [double] (b),
                (b) -- [photon] (c),
                (b) -- [double] (d),
                (d) -- [scalar] (p1),
                (d) -- [scalar] (p2),
                };
            \end{feynman}
        \end{tikzpicture}
        \label{RCT3}
    }
    \quad \quad 
    \subfigure[]{
        \begin{tikzpicture}
            \begin{feynman}
                \vertex [square dot] (a) {};
                \vertex [right= of a,dot] (b){};
                \vertex [above right=1.6cm of b] (p1){\(\pi^0\)};
                \vertex [right= of b] (c) {\(\gamma^*\)};
                \vertex [below right=1.6cm of b] (p2) {\(\pi^{-}\)};
                \diagram* {
                (a) -- [double] (b),
                (b) -- [scalar] (p1),
                (b) -- [photon] (c),
                (b) -- [scalar] (p2),
                };
            \end{feynman}
        \end{tikzpicture}
        \label{RCT4}
    }
    \quad \quad 
    \subfigure[]{
        \begin{tikzpicture}
            \begin{feynman}
                \vertex [square dot] (a) {};
                \vertex [right= of a, dot] (b){};
                \vertex [above right=1.6cm of b] (p1){\(\pi^0\)};
                \vertex [right= of b, dot] (c) {};
                \vertex [right=of c] (d) {\(\gamma^*\)};
                \vertex [below right=1.6cm of b] (p2) {\(\pi^{-}\)};
                \diagram* {
                (a) -- [double] (b),
                (b) -- [scalar] (p1),
                (b) -- [double] (c),
                (c) -- [photon] (d),
                (b) -- [scalar] (p2),
                };
            \end{feynman}
        \end{tikzpicture}
        \label{RCT5}
    }
    \quad \quad 
    \subfigure[]{
        \begin{tikzpicture}
            \begin{feynman}
                \vertex [square dot] (a) {};
                \vertex [above right= of a, dot] (b){};
                \vertex [right= of b] (c){\(\gamma^*\)};
                \vertex [below right= of a, dot] (d){};
                \vertex [above right= of d] (p1) {\(\pi^{0}\)};
                \vertex [right= of d] (p2) {\(\pi^{-}\)};
                \diagram* {
                (a) -- [double] (b),
                (b) -- [photon] (c),
                (a) -- [double] (d),
                (d) -- [scalar] (p1),
                (d) -- [scalar] (p2),
                };
            \end{feynman}
        \end{tikzpicture}
        \label{RCT6}
    }
    \quad \quad
    \subfigure[]{
        \begin{tikzpicture}
            \begin{feynman}
                \vertex [square dot] (a) {};
                \vertex [above right=of a, dot] (b){};
                \vertex [right=of b] (c){\(\gamma^*\)};
                \vertex [right=1.6cm of a] (p1){\(\pi^0\)};
                \vertex [below right=1.6cm of a] (p2){\(\pi^{-}\)};
                \diagram* {
                (a) -- [double] (b),
                (b) -- [photon] (c),
                (a) -- [scalar] (p1),
                (a) -- [scalar] (p2),
                };
            \end{feynman}
        \end{tikzpicture}
        \label{RCT7}
    }
    \quad \quad
    \subfigure[]{
        \begin{tikzpicture}
            \begin{feynman}
                \vertex [square dot] (a) {};
                \vertex [below=of a] (p1) {\(\pi^0\)};
                \vertex [right=of a, dot] (c){};
                \vertex [above right=1.6cm of c] (d) {\(\gamma^{*}\)};
                \vertex [below right=1.6cm of c] (p2) {\(\pi^{-}\)};
                \diagram* {
                (a) -- [scalar] (p1),
                (a) -- [double, edge label'=\(a_{1}^{-}\)] (c),
                (c) -- [photon] (d),
                (c) -- [scalar] (p2),
                };
            \end{feynman}
        \end{tikzpicture}
        \label{RCT8}
    }
\caption{Contributions to  $V^{\mu\nu}$ in RChT  of the hadronic vertex $W^{*-}\to \pi^-\pi^0\gamma^*$ vertices. Those involving resonances are highlighted with a thick dot. Insertion of the weak charged current is represented by the square dot. Resonances ($\rho$ unless specified) are represented with a double line and $\gamma^*$ stands for $\gamma^*\to\ell^+ \ell^-$.}
\label{CFD}
\end{figure}
\end{center}

In the case of a virtual photon, we have contributions from the same Feynman diagrams shown in figure \ref{CFD}, taking due care of $k^2\not =0$. As in the case of a real photon, some of these contributions would appear in the leading order term $V_{\mu\nu}^{LO}$ given in Eq. (\ref{Vmunu}). As it was discussed above, in this case the structure-dependent contributions can be described in terms of seven form factors $v_{1,\cdots ,7}$, defined in Eq. (\ref{VP}) as the coefficients of gauge-invariant structures. An explicit evaluation of them leads to:  
\begin{eqnarray} \label{VUS2}
v_{1} &=& \frac{F_{V}G_{V}}{f^2}\left[ 2D_{\rho}^{-1}\left( k^2 \right) + 2D^{-1}_{\rho}\left( t^{\prime} \right) + tD_{\rho}^{-1}(t)D_{\rho}^{-1}\left(k^2 \right) +
t D_{\rho}^{-1}(t)D_{\rho}^{-1}\left( t^{\prime} \right) \right] \nonumber \\
&\ \ \ \ \ +& \frac{F_{V}^2}{2f^2}\left[ - D_{\rho}^{-1}\left( k^2 \right) - D_{\rho}^{-1}\left( t^{\prime} \right) + \left( t^{\prime} - k^2 \right) D_{\rho}^{-1}\left( t^{\prime} \right)D_{\rho}^{-1}\left( k^2 \right) \right] \nonumber\\
&\ \ \ \ \ +& \frac{F_{A}^2}{f^2 m_{a_1}^2}\left( m_{a_1}^2 - m_{\pi}^2 + \frac{t}{2} \right)D_{a_1}^{-1}\left[ \left( p_{-} + k \right)^2 \right], \nonumber \\
v_{2} &=& \frac{F_{V}G_{V}t}{f^2}\left[ -D_{\rho}^{-1}(t)D_{\rho}^{-1}\left( k^2 \right) - D_{\rho}^{-1}(t)D_{\rho}^{-1}\left( t^{\prime} \right) \right] \nonumber\\
&\ \ \ \ \ +& \frac{F_{V}^2}{2f^2}\left[ -D_{\rho}^{-1}\left( k^2 \right)
- D_{\rho}^{-1}\left( t^{\prime} \right) - \left(t^{\prime} - k^2 \right) D_{\rho}^{-1}\left( t^{\prime} \right)D_{\rho}^{-1}\left( k^2 \right) \right] \nonumber\\
&\ \ \  \ \ +& \frac{F_{A}^2}{f^2m_{a_1}^2}\left( m_{a_1}^2 - m_{\pi}^2 - k\cdot p_{-} \right) D_{a_1}^{-1}\left[ \left( p_{-}+k \right)^2\right] , \nonumber \\
v_{3} &=& \frac{F_{A}^2}{f^2m_{a_1}^2}D_{a_1}^{-1}\left[ \left( p_{-} + k \right)^2 \right],\nonumber  \\
v_{4} &=& \frac{F_{V}^2}{f^2}D_{\rho}^{-1}\left( t^{\prime} \right)D_{\rho}^{-1}\left( k^2 \right) - \frac{2F_{V}G_{V}}{f^2}D_{\rho}^{-1}(t)D_{\rho}^{-1}\left( t^{\prime} \right), \nonumber\\
v_{5} &=& \frac{F_{V}^2}{2f^2}\left[ -D_{\rho}^{-1}\left( k^2 \right) -D_{\rho}^{-1}\left( t^{\prime} \right) - k\cdot \left( p_{0}-p_{-} \right)D_{\rho}^{-1}\left( t^{\prime} \right)D_{\rho}^{-1}\left( k^2 \right)\right] \nonumber \\
& \ \ \ \ \ +& \frac{F_{A}^2}{f^2 m_{a_1}^2}\left( m_{a_1}^2 - m_{\pi}^2 + \frac{t}{2} \right)D_{a_1}^{-1}\left[ \left( p_{-} + k \right)^2 \right], \nonumber \\
v_{6} &=& \frac{2F_{V}G_{V}}{f^2}D_{\rho}^{-1}(t) D_{\rho}^{-1}\left( k^2 \right) + \frac{F_{V}^2}{f^2}D_{\rho}^{-1}\left( t^{\prime} \right)D_{\rho}^{-1}\left( k^2 \right),\nonumber \\
v_{7} &=& \frac{2F_{V}G_{V}}{f^2}D_{\rho}^{-1}(t) D_{\rho}^{-1}\left( k^2 \right) + \frac{F_{V}^2}{f^2}D_{\rho}^{-1}\left( t^{\prime} \right)D_{\rho}^{-1}\left( k^2 \right)+ \frac{F_{A}^2}{f^2m_{a_1}^2}D_{a_1}^{-1}\left[ \left( p_{-} + k \right)^2 \right].
\end{eqnarray}

Note that the above expressions reduce to the corresponding Eqs. (\ref{firstap}) for $v_{1,\cdots ,4}$ in the case of real photons ($k^2\to 0$). Also, we observe that, owing to the conservation of electromagnetic current, the form factors $v_{5,6,7}$ do not contribute to the vector tensor terms in Eq. (\ref{VP}) in the real photon case.

The short-distance constraints for two-point Green functions (which include the set of relations $F_V=\sqrt{2}f$, $G_V=f/\sqrt{2}$, $F_A=f$) get modified when including three-point Green functions in both intrinsic parity sectors \cite{RuizFemenia:2003hm, Cirigliano:2004ue,Cirigliano:2005xn,Cirigliano:2006hb,Mateu:2007tr,Guo:2008sh,Dumm:2009kj,Dumm:2009va,Guo:2010dv,Kampf:2011ty,Dumm:2012vb,Chen:2012vw,Roig:2013baa,Roig:2014uja,Guevara:2016trs,Guevara:2018rhj,Dai:2019lmj, Miranda:2020wdg}. The consistent set of relations in this more general case includes $F_V=\sqrt{3}f$ \cite{Roig:2013baa} that --through the appropriate asymptotic behaviour of the spin-one correlators--  implies, $G_V=f/\sqrt{3}$ and $F_A=\sqrt{2}f$. As studied extensively in ref. \cite{Miranda:2020wdg} for the $\tau^-\to\nu_\tau\pi^-\pi^0\gamma$ decays,  shifting from
\begin{equation}\label{SDconstraints1}
F_V=\sqrt{2}f,\quad G_V=\frac{f}{\sqrt{2}},\quad F_A = f\,,
\end{equation}
to
\begin{equation}\label{SDconstraints2}
F_V=\sqrt{3}f,\quad G_V=\frac{f}{\sqrt{3}},\quad F_A = \sqrt{2}f\,,
\end{equation}
gives a rough estimate of the uncertainty in the calculation with the interaction Lagrangian (\ref{InteractionLagrangian}) due to missing higher-order terms in the chiral expansion. We will take relations (\ref{SDconstraints1}) as the reference ones but evaluate alternatively with (\ref{SDconstraints2}) to assess our model-dependent error.

\noindent 

\subsection{Structure-dependent axial-vector contributions}\label{axc}

The most general form of  the axial-vector weak current contribution  in $\tau^-(P) \to \nu_{\tau}( q ) \pi^-( p_-)\pi^0 ( p_0) \ell^+( p_{\ell^+} ) \ell^- ( p_{\ell^-} ) $ can be built out of the rank-four antisymmetric Levi-Civita tensor and the three independent momenta in the $W^*(W)\to \pi^-(p_-)\pi^0(p_0)\gamma^*(k)$ vertex. Making use of Schouten's identity and the gauge invariance condition $k_{\mu}A^{\mu\nu}=0$, one gets \cite{Bijnens, Cirigliano, Guevara:2016trs} the same result as in the real photon case: 
\begin{eqnarray}\label{Amunu}
A^{\mu\nu} &=& i a_{1}\varepsilon^{\mu\nu\rho\sigma}\left( p_{0}-p_{-} \right)_{\rho}k_{\sigma} + ia_{2}W^{\nu}\epsilon^{\mu\lambda\rho\sigma}p_{-\lambda}p_{0\rho}k_{\sigma} + i a_{3}\varepsilon^{\mu\nu\rho\sigma}k_{\rho}W_{\sigma} \nonumber \\
&\ \ \ \ \ +& i a_{4}\left( p_{0}+k \right)^{\nu}\varepsilon^{\mu\lambda\rho\sigma}p_{-\lambda}p_{0\rho}k_{\sigma},
\end{eqnarray}

\noindent where $W= P-q = p_{-}+p_{0}+k$. As in the vector tensor case, the axial-vector form factors $a_{1\cdots 4}$ are Lorentz invariant functions that depend upon two kinematical Lorentz invariants (in addition to $t'=W^2$ and $k^2$). 

At $O(p^4)$ only $a_{1}$ and $a_{2}$, from the Wess-Zumino-Witten functional (\cite{Wess, Witten}), contribute and are obviously the same as in the real photon case at this order (see Fig. \ref{FDWZW}). They are given by \cite{Cirigliano}
\begin{eqnarray}\label{WZWeq}
a_{1} &=& \left[ 8\pi ^2 f^2 \right]^{-1},\nonumber \\
a_{2} &=& -\left[ 4\pi^2 f^2 \left( t^{\prime} - m_{\pi}^2 \right) \right]^{-1}.
\end{eqnarray}

\begin{center}
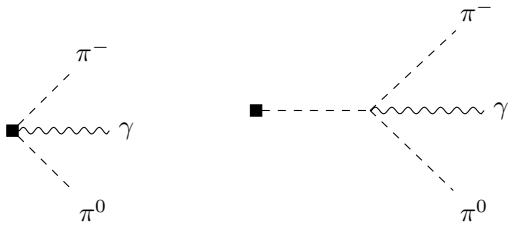
\begin{figure}[h]
    \centering
    \subfigure{
        \begin{tikzpicture}
            \begin{feynman}
                \vertex [square dot] (a) {};
                \vertex [medium, right = of a] (b){\( \gamma \)};
                \vertex [medium, above right = of a] (c) {\( \pi^{-} \)};
                \vertex [medium, below right = of a] (d) {\( \pi^{0} \)};
                \diagram* {
                (a) -- [photon] (b),
                (a) -- [scalar] (c),
                (a) -- [scalar] (d),
                };
            \end{feynman}
        \end{tikzpicture}
        \label{WZW1}
    }
    \quad \quad \quad 
    \subfigure{
        \begin{tikzpicture}
            \begin{feynman}
                \vertex [square dot] (a) {};
                \vertex [right = of a] (b);
                \vertex [right = of b] (c) {\( \gamma \)};
                \vertex [above right = of b] (d) {\(\pi^{-}\)};
                \vertex [below right = of b] (e) {\(\pi^{0}\)};
                \diagram* {
                (a) -- [scalar] (b) -- [photon] (c),
                (b) -- [scalar] (d),
                (b) -- [scalar] (e),
                };
            \end{feynman}
        \end{tikzpicture}
        \label{WZM2}
    }
\caption[]{Contributions from the Wess-Zumino-Witten to the $W^{*-}\to  \pi^{-} \pi ^{0}\gamma^* $ vertex \cite{Wess} \cite{Witten}. The weak charged current is represented by the square dot.}
\label{FDWZW}
\end{figure}
\end{center}

As in the case of radiative $\tau^- \to \nu_{\tau}\pi^- \pi^0  \gamma$ decays, we expect the corresponding contributions to the decay observables in lepton pair production to be negligibly small \cite{FloresTlalpa:2005fz, FloresBaez:2006gf, Miranda:2020wdg}. Therefore, we do not consider necessary to compute all remaining axial-vector contributions which will introduce in addition further (although small) uncertainties in our computation.

\section{Branching ratio and lepton-pair spectrum} \label{BRIMS}

As it is well known, the unpolarized squared amplitude of a five-body decay as $\tau^- \to \nu_{\tau} \pi^-\pi^0\ell^+\ell^-$, depends on eight independent kinematical variables. Depending upon the specific observable we are interested in, it will become necessary to integrate some or all of these kinematical variables. In this paper we find convenient to use the set of invariant variables described in Ref. \cite{Kumar} (see \ref{Kinematics} in which we have defined these variables and have calculated one of the non-trivial scalar products, a subtlety when there are more than four particles in the final state for decay processes).  More specifically, we will compute the invariant mass distribution of the lepton pair ($k^2$-distribution) and the corresponding branching fraction for the tau decays under consideration. The kinematical domain of the lepton pair distribution is the interval $\left[ 4m_{\ell}^2, \left( M_{\tau} - 2m_{\pi} \right) ^2 \right]$, being $\ell = e, \mu$.

In order to distinguish among the different contributions, we split the total decay observables into three terms: 1) the IB piece, {\it i.e.} the inner bremsstrahlung or model-independent contributions obtained with vanishing $v_{1,\cdots ,7}$ and $a_{1,\cdots ,4}$ form factors; 2) the VV (AA) model-dependent part, corresponding to the terms with non-vanishing vector (axial-vector) form factors and, 3) the  IB-V, IB-A and V-A pieces, which correspond to the interferences of IB, V and A contributions.  

Table \ref{BRRes1}  shows the  results of different contributions to the  branching ratios of $e^+e^-$ and $\mu^+\mu^-$ pair production (the numerical errors in the integration are shown within parentheses). These calculations were obtained using the vector form factors given in Eq. (\ref{VUS2}), the axial-vector form factors of Eq. (\ref{WZWeq}) and the short-distance contraints exhibited in Eq. (\ref{SDconstraints1}); for comparison, we also show within square brackets the results obtained using the vector form factors of Eq. (\ref{firstap}), corresponding to the real photon case. In the third (fifth) column of Table \ref{BRRes1} we also show the results for various contributions to the branching fraction obtained using the vector (\ref{VUS2}) and axial-vector (\ref{WZWeq}) form factors, but subject to the short-distance constraints on the couplings constants of resonances shown in Eq. (\ref{SDconstraints2}). As it was explained at the end of Section \ref{sectionvector}, shifting the values of coupling constants according to the prescriptions on short-distance constraints allows us to assess an important part of theoretical uncertainties.

The results shown in Table \ref{BRRes1} exhibit the suppression expected since lepton-pair production is $O(\alpha^2)$ with respect to the dominant $\tau^-\to \pi^-\pi^0\nu_{\tau}$ decay\footnote{%
Noteworthy,  the large inner bremsstrahlung contributions coming from photon emission off the $\tau^-$ lepton and $\pi^-$ meson, is almost cancelled by their interference, 
 which yields  physical results (e. g., branching ratios of order $\alpha^2$). This type of cancellation agrees with that observed in $\tau^{-}\to \nu_{\tau}\pi^{-}\ell^{+}\ell^{-}$ decays \cite{Adolfo}.}. Also, the $\mu^+\mu^-$ pair production is further suppressed with respect to $e^+e^-$ production given that the later is largely dominated by model-independent contributions, which are enhanced and peaked at lower invariant mass values of the lepton-pair invariant mass due to the virtual photon propagator. As pointed out before, the axial-vector contributions are  suppressed in all cases.

\begin{table}[]
\centering
\begin{tabular}{|c||c|c||c|c|}
\hline
& & & & \\
& & $\ell^{+}\ell^{-}=e^{+}e^{-}$ & & $\ell^{+}\ell^{-}=\mu^{+}\mu^{-}$ \\ 
 Contribution & $\ell^{+}\ell^{-}=e^{+}e^{-}$ & using $\eqref{SDconstraints2} $ for  & $\ell^{+}\ell^{-}=\mu^{+}\mu^{-}$ & using $\eqref{SDconstraints2} $ for  \\
& & $F_{V}$, $F_{A}$ and $G_{V}$  & & $F_{V}$, $F_{A}$ and $G_{V}$ \\
& & & & \\
\hline \hline 
& & & & \\
IB  & 2.213(11) $\times$ 10$^{-5}$ &  & 5.961(3) $\times$ 10$^{-8}$ & \\ 
 & \big[ 2.206(11) $\times$ 10$^{-5}$ \big]& &  \big[ 5.958(7) $\times$ 10$^{-8}$ \big]& \\
 & & & & \\
\hline
& & & & \\
VV & 6.745(36) $\times$ 10$^{-7}$ & 9.571(48) $\times$ 10$^{-7}$ & 5.462(4) $\times$ 10$^{-8}$ & 9.429(7) $\times$ 10$^{-8}$ \\ 
 & \big[ 6.442(38) $\times$ 10$^{-7}$ \big] & & \big[ 4.801(3) $\times$ 10$^{-8}$ \big] & \\
& & & & \\
\hline
& & & & \\
AA & 1.91(1) $\times$ 10$^{-8}$ & & 1.663(1) $\times$ 10$^{-9}$ & \\
& \big[ 1.91(1) $\times$ 10$^{-8}$ \big] & & \big[ 1.663(1)$\times$ 10$^{-9}$ \big] & \\
& & & & \\
\hline
& & & & \\
IB-V & $-$3.83(18) $\times$ 10$^{-7}$ & -1.02(18)$\times$ 10$^{-7}$ & 1.337(4) $\times$ 10$^{-8}$ & 2.126(5)$\times$ 10$^{-8}$ \\
& \big[ $-$3.85(18) $\times$ 10$^{-7}$ \big] & & \big[ 5.25(4) $\times$ $10^{-9} \big]$& \\
& & & & \\
\hline
& & & & \\
IB-A & 9.1(4.5) $\times$ 10$^{-9}$ & & 2.85(3) $\times$ 10$^{-9}$ & \\
& \big[ 9.1(4.5) $\times$ 10$^{-9}$ \big] & &\big[ 2.85(3) $\times$ 10$^{-9}$ \big] & \\
& & & & \\
\hline
& & & & \\
V-A & 5.2(2.1) $\times$ 10$^{-9}$ & 4.5(2.6)$\times$ 10$^{-9}$ & $-$1.73(3) $\times$ 10$^{-10}$ & -1.65(5) $\times$ 10$^{-10}$ \\
& \big[ 3.1(2.4) $\times $ 10$^{-9}$ \big]& & \big[ $-$7.9(3) $\times$ 10$^{-11}$ \big]& \\
& & & & \\
\hline
& & & & \\
Total & 2.245(13) $\times$ 10$^{-5}$ & 2.302(13) $\times$ 10$^{-5}$ & 1.319(2) $\times$ 10$^{-7}$ & 1.795(2) $\times$ 10$^{-7}$ \\
& \big[ 2.235(13) $\times$ 10$^{-5}$ \big]& & \big[ 1.173(1) $\times$ 10$^{-7}$ \big]& \\
& & & &\\
\hline\hline
\end{tabular}
\caption{\\ Contributions to the branching ratio of $\tau^- \to \nu_{\tau}\pi^-\pi^0\ell^+\ell^-$ decays.  IB, VV, and AA stand for the Inner Bremsstrahlung, Vector, and Axial-Vector contributions, respectively, while IB-V, IB-A, and V-A correspond to their  interferences. 
Columns three and five display the branching ratios obtained using Eq. (\ref{SDconstraints2}) for the relations of resonance couplings with the pion decay constant entering the structure-dependent vector form factors, while the second and four columns correspond to the use of relations (\ref{SDconstraints1}). }
\label{BRRes1}
\end{table}

Our final predictions for the branching fractions are: 
\begin{eqnarray}\label{FinalResults}
{\rm BR}(\tau^-\to\nu_\tau\pi^-\pi^0e^+e^-)&=&(2.27\pm 0.03)\times 10^{-5},\\ 
{\rm BR}(\tau^-\to\nu_\tau\pi^-\pi^0\mu^+\mu^-)&=&(1.55\pm 0.25)\times 10^{-7}\ .
\end{eqnarray}
The associated errors  cover the results shown  in the different columns of Table \ref{BRRes1}\footnote{We note that these errors are larger than $1/N_C$, typical of a large-$N_C$ expansion, for the structure-dependent contributions. }.

\begin{figure}[h]
\centering
\includegraphics[width=15cm]{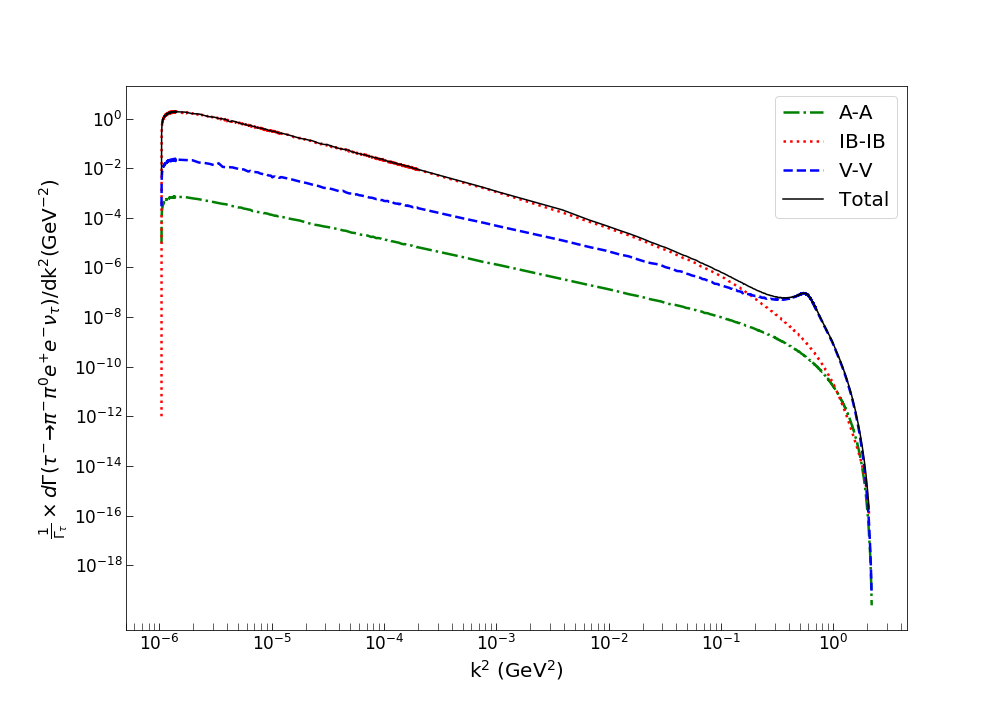}
  \caption{Contributions to the normalized invariant mass distribution for  $e^{+}e^{-}$ pair production (interferences are not displayed). 
   A double logarithmic scale was used. 
  The second peak is due to the $\rho(770)$ dominance of the virtual photon propagator.}
  \label{Totee_sep}
\end{figure}


\begin{figure}[h]
\centering
\includegraphics[width=15cm]{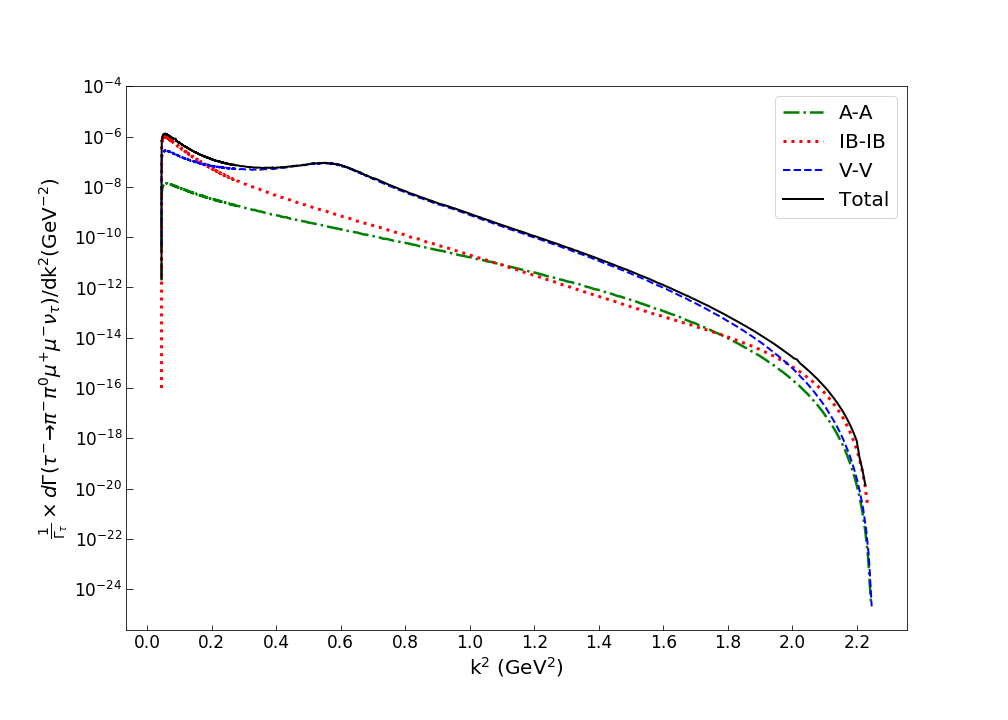}
  \caption{Contributions to the normalized invariant mass distribution for $\mu^{+}\mu^{-}$ pair production (inteferences are not displayed). 
 The second peak is due to the  $\rho(770)$ dominance of the virtual photon propagator. 
  }
  \label{Totmumu_sep}
\end{figure}




The normalized (to the total $\tau$ decay width) lepton-pair invariant mass distributions are shown in Figure \ref{Totee_sep} for electron-positron and in Figure \ref{Totmumu_sep} for $\mu^+\mu^-$ production. Both distributions are peaked very close to the corresponding threshold ($k^2_{\rm thr}=4m^2_{\ell}$) for lepton pair production, with an enhanced peaking for $e^+e^-$ production,  due to the $1/k^4$ dependence of the squared amplitude. The second peak in the plots corresponds to the $\rho^0(770)-\gamma^*$ couplings dominance in the vector form factors. It is also clear that the model-dependent contributions are more visible in the $\mu^+\mu^-$ than in $e^+e^-$ production, which is also related to the suppression of inner bremsstrahlung for large photon virtualities.

\section{Conclusions} \label{Conclusions}
We have calculated for the first time the branching ratios and lepton-pair mass distributions of the five-body decays $\tau^{-}\to \nu_{\tau} \pi^{-}\pi^{0}\ell^{+}\ell^{-}$ 
($\ell=e,\mu$)
. As expected, these observables are of $O(\alpha^2)$ with respect to the corresponding dominant di-pion $\tau$ lepton decay. For the $\ell=e$ case
, a clear inner bremsstrahlung (IB) dominance is observed due to the small $\ell^+\ell^-$ invariant mass ($k^2$) threshold values. 
 On the other hand, 
for $\ell=\mu$, both contributions, structure-dependent and IB, are of the same order.

The structure-dependent contributions corresponding to the $W^{-}\to \pi^{-}\pi^{0}\gamma^{*}$ effective vertex, were calculated using the Resonance Chiral Theory framework
. Such an approach considers the lightest resonances as active degrees of freedom giving the low-energy chiral limit of 
QCD and ensuring an appropriate short-distance behaviour. The structure-dependent vector form factors coincide (in the limit $k^2\to 0$) with their counterparts computed in the case of the radiative $\tau^- \to \nu_{\tau}\pi^-\pi^0\gamma$ decays \cite{Cirigliano}. We expect axial-vector structure-dependent contributions to be negligible and we stick to their values provided by the Wess-Zumino-Witten anomalous terms.

Within this framework, we get $\mathcal{BR}\left(\tau^{-}\to \nu_{\tau}\pi^{-}\pi^{0} e^{+}e^{-}\right) = 2.27(3)\times 10^{-5}$ (which is essentially free of hadronic uncertainties) and $\mathcal{BR}\left(\tau^{-}\to \nu_{\tau}\pi^{-}\pi^{0}\mu^{+}\mu^{-} \right) = 1.55(25)\times 10^{-7}$. The estimated theoretical uncertainties are associated to different relations  between resonances couplings and the pion decay constant, obtained from  the short-distance behavior of two- and three-point Green functions. 
While the branching fraction for $e^+e^-$  channel allows to conclude that it could be discovered already with BaBar or Belle data, the $\mu^+\mu^-$ case will challenge the capabilities of Belle-II. On the other hand, the measurement of the $\mu^+\mu^-$  spectrum, which is more sensitive to structure-dependent contributions, can be useful to test previous calculations of radiative corrections to di-pion tau lepton decays. Therefore, it has the potential of reducing the uncertainties on the dominant piece of the hadronic vacuum polarization part of $a_\mu$ using tau data.

Finally, the addition of the 
matrix elements derived in this work to the Monte Carlo generator  TAUOLA \cite{Jadach:1993hs} will be useful in improving background rejection for searches of three-prong lepton flavor or lepton number violating tau decays. 

\section*{Acknowledgements}
J.~L.~G.~S. thanks Conacyt for his Ph. D. scholarship.  G.~L.~C. received support from Ciencia de Frontera Conacyt project  No. 428218. P.~R.~ is indebted for the funding received through Fondo SEP-Cinvestav 2018 (project No. 142) and C\'atedra Marcos Moshinsky (2020).

\appendix

\section{Five-body kinematics}\label{Kinematics}

The kinematics of the five-body decay process $\tau^{-}(P)\to \nu_{\tau}(q) \pi^{-}(p_{-}) \pi^{0}(p_{0}) \ell^{+}(p_{\ell^{+}})\ell^{-}(p_{\ell^{-}})$ where the lepton pair $\ell^{+}\ell^{-}$ is either $e^{+}e^{-}$ or $\mu^{+}\mu^{-}$ and $k = \ell^{+}+\ell^{-}$ 
is the momentum of the virtual photon, is described in terms of eight
independent variables. All the  scalar products of two-momenta can be written in terms of these independent kinematical variables.  Following reference \cite{Kumar} we choose these 
variables as follows:
\begin{align*} 
s_{1}&=(P-q)^2,   &   s_{2}&=(P-q-p_{-})^2,    &    s_{3}&=(P-q-p_{-}-p_{0})^2,\nonumber\\
u_{1}&=(P-p_{-})^2,   &    u_{2}&=(P-p_{0})^2,   &    u_{3}&=(P-p_{\ell^{-}})^2,\nonumber\\
t_{2}&=(P-p_{-}-p_{0})^2,   &    t_{3}&=(P-p_{-}-p_{0}-p_{\ell^{-}})^2,
\label{ts}
\end{align*}

 \noindent and the auxiliary variables $s_{0}=M_{\tau}^2$, $s_{4}=m_{\ell^{+}}^2$, $u_{0}=s_{1}$ and $t_{1}=u_{1}$. 

In general, for decay processes with $n$ particles in the final state, it can be shown that we will have $3n-7$ independent invariants. 
 In the case $n\ge 5$, it is well known that some scalar products 
cannot be written directly in terms of the  $s_{i}$, $t_{i}$ and $u_{i}$ variables. This is the case for the  $p_{-}\cdot p_{\ell^+}$ and $p_{0}\cdot p_{\ell^+}$ scalar products. 
Following reference \cite{Kumar}
and making use of symmetry considerations, $p_{-}\cdot p_{\ell^+}$ reads
\begin{eqnarray}\label{eqn4}
p_{-} \cdot p_{\ell^+}=a\left(P\cdot p_{-} - q\cdot p_{-}- p_{-}^2 - p_{-}\cdot p_{0} \right)+b\left( P\cdot p_{-} - p_{-}^2-p_{-}\cdot p_{0}\right)+c \left(P\cdot p_{-}\right).
\end{eqnarray}


\noindent with $a$, $b$ and $c$ given as follows,
\begin{eqnarray}
a &=& \frac{1}{Z}\left[-ADF+AEs - BDE+ BFt_{2}+C\left(D^2-M_{\tau}^2 t_{2}\right)\right],
\nonumber\\  
b &=& \frac{1}{Z}\left[-B(AF+CD)+M_{\tau}^2AC+B^2E+DFs_{3}-M_{\tau}^2Es_{3}\right],
\nonumber\\
c &=& \frac{1}{Z}\left[A^2 F-A(BE+CD)+BCt_{2}+DEs_{3}-Fs_{3}t_{2}\right],\nonumber
\end{eqnarray}

\noindent where the capital letters are defined in terms of the already known scalar products in the following way:


\begin{eqnarray}
A &=& \left( P-q-p_{-}-p_{0} \right)\cdot \left( P-p_{-}-p_{0} \right), \nonumber\\
B &=& P \cdot \left(P-q-p_{-}-p_{0} \right),\nonumber\\
C &=& p_{\ell^+}\cdot \left(P-q-p_{-}-p_{0}\right),\nonumber\\
D &=& P\cdot \left( P-p_{-}-p_{0} \right),\nonumber\\
E &=& p_{\ell^+}\cdot \left( P-p_{-}-p_{0} \right),\nonumber\\
F &=& P\cdot p_{\ell^+},\nonumber\\
Z &=& M_{\tau}^2 A^2-2 ABD+t_{2}\left( B^2-M_{\tau}^2s_{3} \right)+D^2 s_{3}.\nonumber
\end{eqnarray}

Then, once we have calculated $p_{-}\cdot p_{\ell^+}$ it is straighforward to obtain (a very lenghty expression for) $p_{0}\cdot p_{\ell^+}$, which we do not quote.

\bibliographystyle{ieeetr}
\bibliography{bibfile}


\end{document}